\documentclass[conference, 10pt]{IEEEtran}
\IEEEoverridecommandlockouts
\usepackage{cite}
\usepackage{amsmath,amssymb,amsfonts,bm}
\usepackage{algorithmic}
\usepackage{multirow}
\usepackage{graphicx}
\usepackage{textcomp}
\usepackage{xcolor}
\def\BibTeX{{\rm B\kern-.05em{\sc i\kern-.025em b}\kern-.08em
    T\kern-.1667em\lower.7ex\hbox{E}\kern-.125emX}}

\usepackage{siunitx}
\usepackage{pgfplots,tikz}
    \usepgfplotslibrary{units} 
	\pgfplotsset{compat=1.9}
	\usepgfplotslibrary{groupplots}
	\usetikzlibrary{positioning}
\makeatletter
\pgfplotsset{
 unit code/.code 2 args=
   \begingroup
   \protected@edef\x{\endgroup\si{#2}}\x
}
\usetikzlibrary{external}
\usepackage{subfig}
\usepackage{commath}
\usepackage[colorlinks=true,linkcolor=black,citecolor=black,urlcolor=black]{hyperref}

\newcommand{\nmse}{\operatorname{NMSE}}
\newcommand{\ncc}{\operatorname{NCC}}

\begin{document}
\title{Near field Acoustic Holography on arbitrary shapes using Convolutional Neural Network}

\author{
\IEEEauthorblockN{Marco Olivieri, Mirco Pezzoli, Fabio Antonacci, Augusto Sarti}
\IEEEauthorblockA{\textit{Dipartimento di Elettronica, Informazione e Bioingegneria (DEIB), Politecnico di Milano}\\
Piazza Leonardo Da Vinci 32, 20133 Milan, Italy
\\
marco1.olivieri@polimi.it, mirco.pezzoli@polimi.it,
fabio.antonacci@polimi.it, augusto.sarti@polimi.it}
}


\maketitle

\begin{abstract}

Near-field Acoustic Holography (NAH) is a well-known problem aimed at estimating the vibrational velocity field of a structure by means of acoustic measurements.
In this paper, we propose a NAH technique based on Convolutional Neural Network (CNN).
The devised CNN predicts the vibrational field on the surface of arbitrary shaped plates (violin plates) with orthotropic material properties from a limited number of measurements.
In particular, the architecture, named Super Resolution CNN (SRCNN), is able to estimate the vibrational field with a higher spatial resolution compared to the input pressure.
The pressure and velocity datasets have been generated through Finite Element Method simulations.
We validate the proposed method by comparing the estimates with the synthesized ground truth and with a state-of-the-art technique.
Moreover, we evaluate the robustness of the devised network against noisy input data.

\end{abstract}


\section{Introduction}
Near-field acoustic holography (NAH) \cite{maynard1985nearfield, williams1999fourier} is an interesting acoustic-based technique for the contactless analysis of vibrating structures such as plates and shells. 
NAH represents an appealing alternative to vibrational analysis carried out with accelerometric sensors when, for example, the structure under analysis is particularly fragile or the deployment of accelerometers is not feasible. 
Contactless analysis is also preferred when lightweight objects are considered, since no additional mass needs to be added.
Differently from contactless optical techniques, e.g., Laser Doppler Vibrometer (LDV), NAH can be employed with objects made of reflective materials.

NAH estimates the velocity field of a vibrating structure starting from acoustic measurements acquired in its proximity.
The acoustic pressure is typically captured by a microphone array deployed on a plane, known as holographic plane.
The sensors are placed close to the vibrating surface in order to retrieve the evanescent wave components \cite{williams1999fourier}.
With the aim of estimating the velocity field of the source from the pressure on the holographic plane, NAH relies on the inversion of the well-known Kirchhoff-Helmholtz (KH) integral \cite{williams1999fourier, norton_karczub_2003}.
This operation is known to be a highly ill-conditioned problem, thus many regularization strategies for NAH have been proposed in the literature \cite{williams2001regularization, kim2004optimal, scholte2008wavenumber}.

A direct approach to NAH considers the discretization of the KH integral, which leads to the Boundary Element Analysis (BEA)\cite{banerjee1981boundary, cheng2005bem}, adopted to solve the forward problem. 
Therefore, NAH is implemented through the inversion of BEA (IBEA) \cite{veronesi_digital_1989,schuhmacher2003ibem}, using Tikhonov regularization. 
This technique is able to provide accurate results, but it is severely limited by the extreme computational cost.

An alternative regularization strategy is represented by Compressed Sensing (CS) in \cite{chardon2012near,nachosURL}, where the solution to NAH is approximated by a sparse set of plane wave components.
However, the use of CS is limited to star-shaped planar plates.

A different approach to NAH is given by the Equivalent Source Method (ESM) \cite{koopman1989wavesuperposition,lee2016esmreview}.
This model assumes that the measured acoustic pressure field radiated by the source can be equivalently expressed as the sound field generated by a set of point-like virtual sources located within or in proximity of the real source.
The main problem of ESM is the computation of the optimal set (in terms of number and location) of equivalent sources. 
In order to deal with this problem, ESM techniques based on CS \cite{fernandez2017sparse,ANTONI2019289, canclini2017dictionary} have been proposed with the aim of finding small and sparse subsets of equivalent sources. 

In \cite{canclini2017dictionary} a dictionary-based ESM (DESM) is proposed in order to consider a sparse domain for solving ESM limitations. 
The ESM solution space is restricted to a suitable compressed dictionary whose components are retrieved from several sets of equivalent sources. 
The dictionary is built from synthetic data varying the mechanical parameters of the object, while the object dimensions are fixed and known. 
The resulting set of equivalent sources weights is reduced using principal component analysis and then it constitutes the learned dictionary. 
Nevertheless, the location and the number of equivalent sources used to build the dictionary are still an open problem. 
This is especially true when the geometry of the objects under study are complex, e.g., whose surface exhibit curvatures. 

Recently, a new approach to NAH based on Deep Learning \cite{Goodfellow2016Deeplearing} has been proposed in \cite{olivieri2020nah}. 
The authors, inspired by the effectiveness of learned features for NAH \cite{canclini2017dictionary} and the well-known feature learning capabilities of Deep Neural Networks (DNN) \cite{bainco2019acousticdeepreview, he2016deep, campagnoli2020vibrational, acerbi2021interpolation},  proposed a Convolutional Neural Network (CNN) \cite{rawat2017cnnimagereview} for performing NAH. 
The promising approach of \cite{olivieri2020nah} provides accurate results but the evaluation is limited to rectangular plates of isotropic material only.
Moreover, it considers a dense spatial sampling of the hologram at a minimal distance from the source, limiting the adoption in practical scenarios.


In this paper we propose an enhanced version of \cite{olivieri2020nah} with a deep learning solution to NAH based on CNN with UNet \cite{ronneberger2015unet} structure. 
The goal of this work is twofold.
On the one side, we extend the analysis of the CNN-based NAH from the isotropic rectangular plates to 3D structures with complex shapes and curvatures made of orthotropic material, i.e., violin plates.
On the other side, unlike the high spatial sampling requirement needed in the previous work, we decrease the number of microphones in the array, thus having less pressure measurement points on the hologram plane.
Nevertheless, we retain the same velocity spatial resolution of \cite{olivieri2020nah} on the CNN output.
In order to increase the output spatial resolution with respect to the input data, we introduce a super-resolution section in the network structure.
Therefore we refer to the proposed method as super resolution CNN-based NAH (SRCNN-NAH).
Moreover, we also relax the constraint on the distance between holographic plane and object, working now at distances compatible with practical scenarios.
The proposed method is validated comparing the predicted vibrational fields with the ground truth and estimates given by DESM \cite{canclini2017dictionary}.
Simulation results confirm the effectiveness of the proposed super resolution CNN approach to NAH also in the presence of noisy input data.

The paper in structured as follows.
Sec.~\ref{sec:dataModel} presents the data model adopted. In Sec.~\ref{sec:network} the description of the proposed CNN is presented. The methodology and the validation results are reported in Sec.~\ref{sec:results}. Finally, Sec.~\ref{sec:conclusion} draws some final conclusions.
\section{Data model and problem formulation}\label{sec:dataModel}
Let us consider a vibrating object occupying a volume $\mathcal{V}$ with an arbitrarily shaped surface $\mathcal{S}$.

We denote $\bm{s}=[x',y',z']^T$ a point on $\mathcal{S}$ and $\bm{n}$ the outward normal direction unit vector.
The sound pressure at a point $\bm{r}=[x,y,z]^T$  placed outside the volume $\mathcal{V}$ is found by means of the Kirchhoff-Helmholtz (KH) integral as \cite{koopman1989wavesuperposition}
\begin{equation} \label{eq:kirchhoffhelmholtz}
\begin{aligned}
    p(\bm{r}, \omega) &= \int_\mathcal{S} p(\bm{s}, \omega) \frac{\partial}{\partial \bm{n}} g_{\omega}(\bm{r},\bm{s}) d\bm{s} \\ 
    &- j \omega \rho_0 \int_\mathcal{S} v_{n}(\bm{s},\omega) g_{\omega}(\bm{r},\bm{s}) d\bm{s},
\end{aligned}
\end{equation}
where $p(\bm{s},\omega)$ and $v_{\mathrm{n}}(\bm{s},\omega)$ are the sound pressure and the normal velocity evaluated in each point belonging to the surface $\mathcal{S}$, $\omega$ is the angular frequency and $\rho_0 \approx$ \SI{1.2}{\kilogram\per\metre\cubed} is the air mass density at \SI{20}{\celsius}.
The term
\begin{equation} \label{eq:greenfunction}
    g_{\omega}(\bm{r},\bm{s}) = \frac{1}{4\pi}\frac{e^{-j\frac{\omega}{c}\norm[0]{\bm{r}-\bm{s}}}}{\norm[0]{\bm{r}-\bm{s}}}
\end{equation}
represents the Green's function \cite{williams1999fourier}, which models the acoustic wave propagation in the free-field from point $\bm{s}$ to point $\bm{r}$, with $c$ being the sound speed in air and $j$ the imaginary unit.

Let us consider $M$ measurement positions $\bm{r}_1,\ldots,\bm{r}_M$ lying on the plane $\mathcal{H}$, placed in the proximity of the vibrating surface, called as holographic plane.
Moreover, consider a sampled version of the normal velocity field evaluated at the points $\bm{s}_1,\ldots,\bm{s}_N$ lying on the surface $\mathcal{S}$.
We can express the descretized form of \eqref{eq:kirchhoffhelmholtz} with the introduction of the discrete estimator $\mathcal{F}$ as
{\small\begin{equation} \label{eq:discretekirchhoffhelmholtz}
    \mathbf{p}_{\mathcal{H}}(\omega) \approx \mathcal{F}(\mathbf{p}_{\mathcal{S}}, 
    \mathbf{v}, \omega) = \mathbf{G}_p^H(\omega)\mathbf{p}_{\mathcal{S}}(\omega) - j \omega \rho_0 \mathbf{G}_v^H(\omega)\mathbf{v}(\omega), 
\end{equation}}
where $\mathbf{p}_{\mathcal{H}} \in \mathbb{C}^{M\times 1}$ and $\mathbf{p}_{\mathcal{S}} \in \mathbb{C}^{N\times 1}$ are the pressure column vectors evaluated on the holographic plane and on the surface $\mathcal{S}$, respectively, $\mathbf{G}_v \in \mathbb{C}^{N\times M}$ is the matrix of Green's functions which relates the $N$ points on the surface with the $M$ points on the hologram and $\mathbf{G}_p = \frac{\partial}{\partial \bm{n}} \mathbf{G}_v$, the term $\mathbf{v} \in \mathbb{C}^{N\times 1}$ refers to the sampled velocity field on the vibrating surface.
Notice that the number of points $M$ and $N$ can be different and in particular $M<N$, thus having fewer pressure measurements than  velocity estimates.

In the context of NAH, we aim at reconstructing the normal velocity $\mathbf{v}(\omega)$
starting from the pressure measurements acquired at the hologram, denoted with the symbol $\mathbf{p}_{\mathcal{H}}(\omega)$.
In practice, we are interested to the inversion of \eqref{eq:discretekirchhoffhelmholtz}, namely
\begin{equation} \label{eq:inverse_helmholtz}
    \hat{\mathbf{v}}(\omega) \approx \mathcal{F}^{-1}(\mathbf{p}_{\mathcal{H}}(\omega)),
\end{equation}
where $\hat{\mathbf{v}}(\omega)$ is the estimate of the normal velocity field.
However, \eqref{eq:inverse_helmholtz} is a highly ill-posed problem \cite{williams2001regularization}, thus it requires a regularization procedure.

Similarly to \cite{olivieri2020nah}, we propose to approximate the inverse operator $\mathcal{F}^{-1}$ for the velocity magnitude in \eqref{eq:inverse_helmholtz} by means of a CNN.
We present an enhanced version of the network described in \cite{olivieri2020nah} in order to estimate the vibrational field on complex surfaces starting from pressure measurements with low spatial resolution, i.e., employing a lower number of microphones.
\section{Network description}
\label{sec:network}
In this section the details of the proposed model for the super resolution NAH problem (SRCNN-NAH) will be presented.
In particular, the architecture consists of a modified version of the UNet \cite{ronneberger2015unet} structure adopted in \cite{olivieri2020nah} with the addition of a super-resolution section.
The additional section allows us to deal with a higher resolution of the output with respect to the input.
Moreover, a modified loss function enables the analysis of arbitrarily shaped objects introducing binary masks that select the points belonging to each object.

The devised SRCNN-NAH network can be split into four parts: encoder ($\mathcal{E}$), bottleneck ($\bm{\beta}$), decoder ($\mathcal{D}$), and super-resolution section ($\mathcal{SR}$).
The first component $\mathcal{E}$ is responsible to extract a feature map from the input data. This compressed representation is encoded in the bottleneck $\bm{\beta}$ and expanded by the decoder $\mathcal{D}$ .
Finally, the super-resolution $\mathcal{SR}$ section provides the upsampling at the desired output dimensions.
\subsection{Input/Output data}\label{subsec:data}
In order to exploit the spatial relation among the sampled points, we rearranged the pressure and velocity vectors in matrices with dimensions $M=M_1\times M_2$ and $N=N_1\times N_2$, respectively. In particular, we consider the magnitude of each field, yielding
\begin{equation}\label{eq:matrics}
    \mathbf{P}_{\mathcal{H}}(\omega) \in \mathbb{R}^{M_1\times M_2} \quad \text{and} \quad
    \mathbf{V}(\omega) \in \mathbb{R}^{N_1\times N_2},
\end{equation}
where $\mathbf{P}_{\mathcal{H}}(\omega)$ is the magnitude pressure matrix sampled at the hologram plane and $\mathbf{V}(\omega)$ is the magnitude of the normal velocity on the surface.
For the ease of the reader, we will refer to them as pressure and velocity images, respectively.

Let $\mathbf{X}$ be the input dataset of the network collecting all the pressure images with dimensions $M_1 = M_2 = 8$ evaluated at the hologram plane.
Conversely, the dataset $\mathbf{Y}$ consists of the desired output velocity images with $N_1=16$ and $N_2=64$ evaluated in the $N = 1024$ points on the vibrating surface.
Moreover, each image of the dataset has been normalized so that its values are within the range $[0, 1]$.

Furthermore, in order to take into account the object shape let be $\mathbf{B}$ the dataset collecting the binary masks $B\in \mathbb{R}^{N_1\times N_2}$ that select only the points belonging to the surface.
In particular, $[B]_{n_1, n_2} = 1$ if the $(n_1,n_2)$th point is on the structure and $[B]_{n_1, n_2} = 0$, otherwise.
\begin{figure}[tb]
    \centering
    \includegraphics[width=0.47\textwidth]{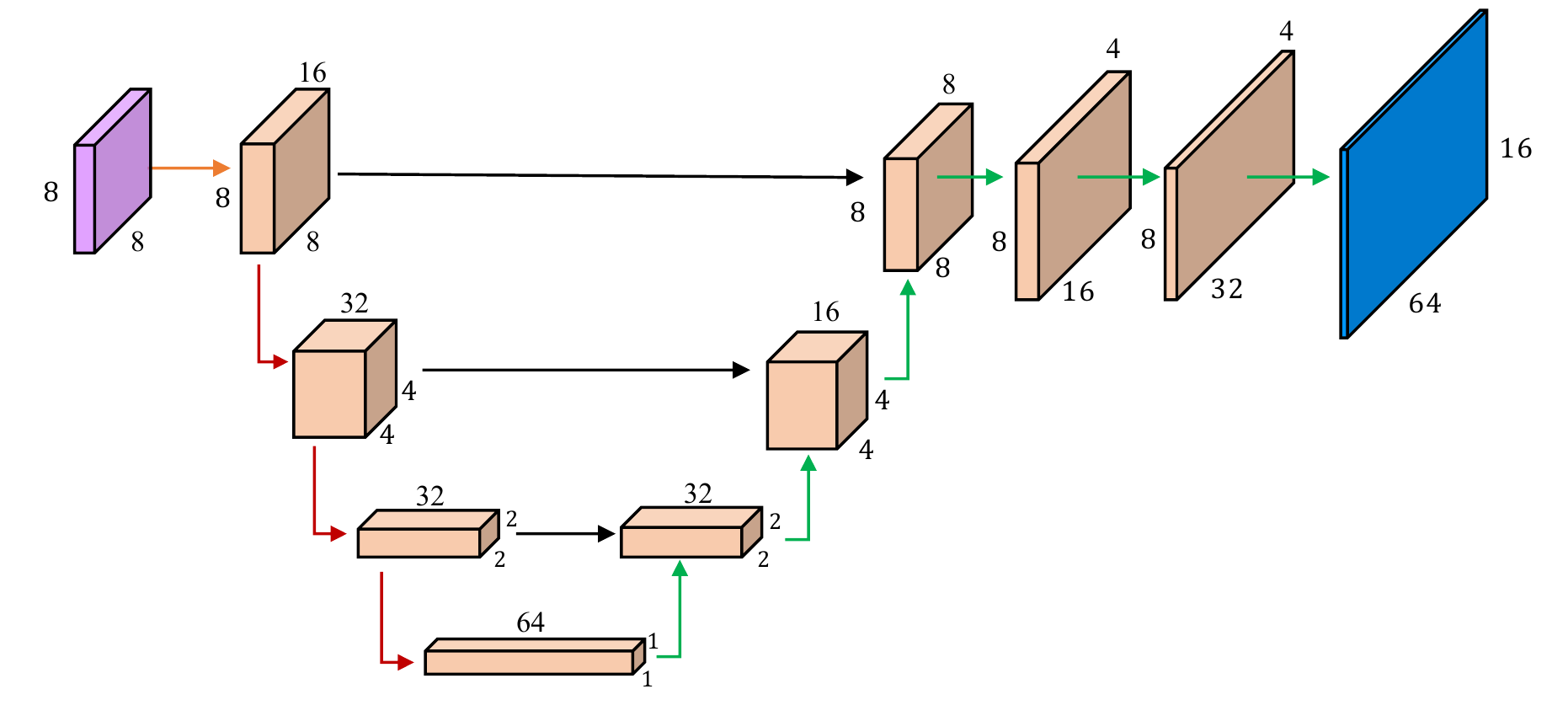}
    \vspace*{-1.5mm}
    \caption{Block diagram of the proposed CNN architecture.} \label{fig:cnn}
\end{figure}

\subsection{Architecture layers}\label{subsec:layers}
In Fig.~\ref{fig:cnn} the detailed structure of the CNN architecture is shown along with the description of the layers dimensions.
After some preliminary evaluations, we resorted on these specific parameters.
The proposed encoder $\mathcal{E}$ is composed of four layers. 
Each one consists of two consecutive layers of 2D convolutions with the same number of filters, kernel size $3\times3$ and ReLU activation functions \cite{nair2010rectified} followed by Batch Normalization.
The compression is achieved with $2\times2$ max pooling operations.
The number of filters follows this sequence: i) $16$, ii) $32$, iii) $32$, iv) $64$, thus the innermost layer $\bm{\beta}$ has dimensions $1\times1\times64$.

The decoder $\mathcal{D}$ reverses the encoder architecture taking as input the bottleneck embedding and performing up-convolutions with stride $2\times2$ with the following number of filters: v) $32$, vi) $16$, vii) $8$. 
The skip connections \cite{he2016deep} are obtained by concatenating pairwise the encoder layers with the corresponding layers of the decoder (see Fig.~\ref{fig:cnn}).

Furthermore, the super-resolution section ($\mathcal{SR}$) consists of two up-convolution blocks with asymmetric strides $1\times2$ followed by ReLU and Batch Normalization, and a last layer with strides $2\times2$. 
This allows us to reach the desired dimensions $N_1\times N_2 = 16\times64$ at the output. 

\subsection{Training Procedure}\label{subsec:training}
The relation between the estimated velocity magnitude and the corresponding sound pressure field defined in \eqref{eq:inverse_helmholtz} can be expressed in terms of the proposed architecture as
\begin{equation}\label{eq:networkpretiction}
    \mathbf{\hat{V}}(\omega) \approx
    \mathcal{D}\left(\mathcal{E}(\mathbf{P}_{\mathcal{H}}(\omega);\mathbf{w}\right);\mathbf{w})\odot B,
\end{equation}
where $\mathbf{\hat{V}} \in \mathbb{R}^{N_1\times N_2}$ is the estimate image of the normal velocity field, $B \in \mathbf{B}$ is the binary mask applied to the network output in order to consider only the points belonging to the vibrating object, $\odot$ is the Hadamard product and $\mathbf{w}$ are the parameters learned by the network.

In order to find the optimal parameters $\mathbf{\hat{w}}$, we trained the devised architecture to minimise the following Mean Square Error (MSE) loss function 
\begin{equation}\label{eq:lossfunction}
    \mathcal{L} = \norm[1]{\mathbf{Y}-\mathcal{D}\left(\mathcal{E}(\mathbf{X};\mathbf{w}\right);\mathbf{w}) \odot \mathbf{B}}^2_2,
\end{equation}
with the default parameters of Adam optimizer presented in \cite{kingma2014adam}.
Notice that the use of $\mathbf{B}$ forces the network to compute the loss only on the structure points.

The network is implemented\footnote{\url{https://github.com/polimi-ispl/nah-srcnn}} in Python using Keras \cite{chollet2015keras} and it was trained for $100$ epochs decreasing the learning rate by a factor $0.2$ on learning plateau. 
Early Stopping regularization technique is also adopted to stop the training after $20$ epochs in which no improvement of the validation loss is shown, thus preventing overfitting.
\section{Results}\label{sec:results}
\subsection{Simulation Setup}\label{subsec:simulation}
In order to extend the CNN-based NAH approach to shapes different from the rectangular plates adopted in \cite{olivieri2020nah}, we decided to work with violin plates with variable outline.

The dataset is composed of $1568$ different synthetic meshes of violin top plates with constant thickness and arching generated by the parametric model introduced in \cite{gonzalez2021data, salvi2021parametric}.
The $20$ parameters defining the shapes are varied according to Gaussian distributions centered around the parameters of a reference violin, as described by the authors in \cite{gonzalez2021data, salvi2021parametric}.
Notice that the use of geometries with parametric outlines eases a generalization on the shapes.
Therefore, we computed the vibration and the radiated acoustic pressure of the plates through finite element analysis using \textit{COMSOL Multiphysics}\textsuperscript{\sffamily\textregistered} \cite{comsol2018v54}. As regards the material, we adopted the values referred to Sitka spruce in \cite{ross2010wood}.

We excited each plate at its eigenfrequencies $\Bar{\omega}$ in the range $[0, \omega_{\mathrm{MAX}}]$ where $\frac{\omega_{\mathrm{MAX}}}{2\pi}=\SI{2000}{\hertz}$. 
The acoustic pressure is sampled on a uniform rectangular grid with dimensions $M_1\times M_2 = 8\times8$ and with spatial sampling steps defined in order to entirely contain the projection of the object on the hologram plane.
As regards the vibration normal velocity, we considered a rectangular grid of $N_1\times N_2 = 16\times64$ points spanning the same dimension of the relative pressure images.
Then, the vibrational data have been post-processed to retrieve the set $\mathbf{B}$ of binary masks needed for the training phase \eqref{eq:lossfunction}.
Differently from \cite{olivieri2020nah}, we increased the distance of the hologram from the source to $\SI{2}{\centi\meter}$, where the distance value is taken from the point of maximum elevation of the violin plate.
Therefore, the input and output datasets used for the proposed SRCNN-NAH approach collect the magnitude of $D = \SI{72323}{}$ different pressure and normal velocity fields, each normalized in the range $[0, 1]$, namely
\begin{equation}\label{eq:dataset}
\begin{aligned}
    \mathbf{X} &\in \mathbb{R}^{M_1\times M_2\times D}, \quad
    \mathbf{Y} \in \mathbb{R}^{N_1\times N_2\times D}\\
    &\quad \text{and} \quad \mathbf{B} \in \mathbb{R}^{N_1\times N_2\times D}.
\end{aligned}
\end{equation}

The dataset was splitted in $\SI{80}{\percent}$ for the training set, $\SI{10}{\percent}$ for the validation set and $\SI{10}{\percent}$ for the test set.
The latter has been built in order to represent $\SI{156}{}$ different violins randomly selected.
Notice that this fact results in testing the network on objects with shapes and dimensions unseen in the training phase.
\begin{figure}[tb]
    \centering
    \includegraphics[width=0.47\textwidth]{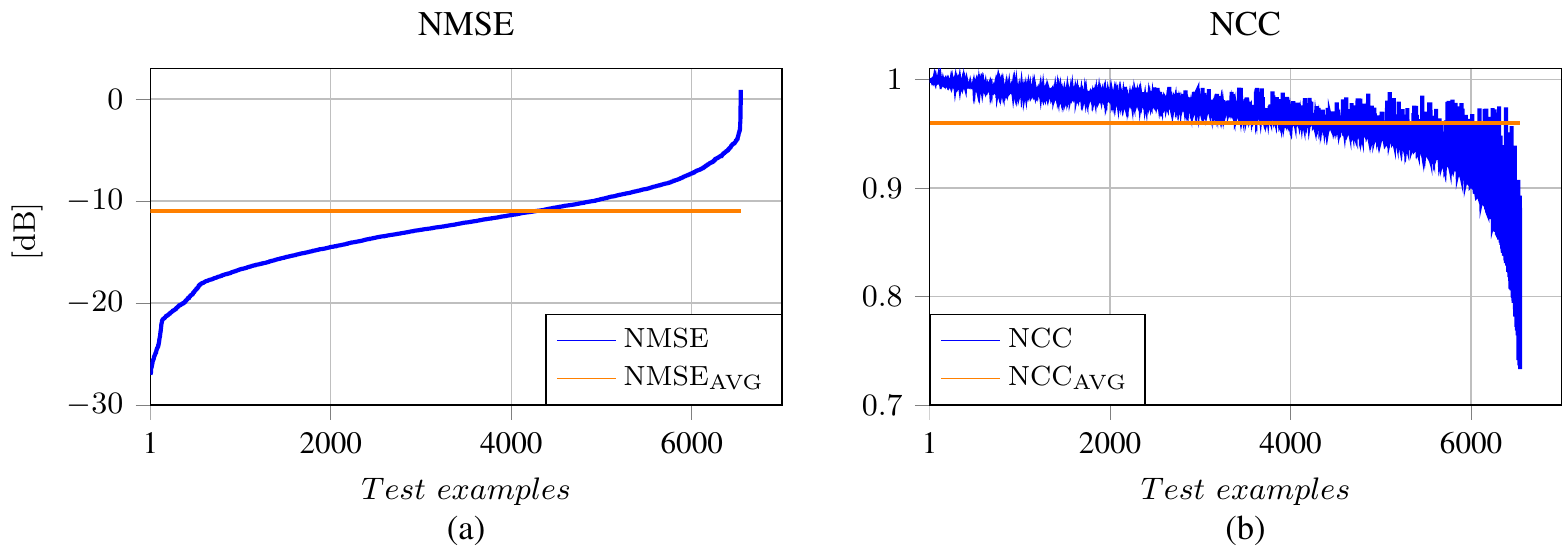}
    \vspace*{-1.5mm}
    \caption{The ordered metrics evaluated on the synthesised test set: (a) NMSE and (b) NCC along with its average values.}
    \label{fig:metrics_cnn}
\end{figure}
\subsection{Metrics Description}\label{subsec:metrics}
The performance of the proposed network is assessed by comparing the estimated vibrational field $\mathbf{\hat{v}}(\omega)$ \eqref{eq:inverse_helmholtz}, given by SRCNN-NAH, with the simulated ground truth $\mathbf{v}(\omega)$ using two different metrics.
The Normalized Cross Correlation ($\ncc$) defined as
\begin{equation}\label{eq:ncc}
    \ncc \left( 
    \mathbf{\hat{v}}(\omega), \mathbf{v}(\omega)\right) = 
    \frac{|\mathbf{\hat{v}}(\omega)^T \mathbf{v}(\omega)|}
    {\norm[1]{\mathbf{\hat{v}}(\omega)}_2 \cdot \norm[1]{\mathbf{v}(\omega)}_2},
\end{equation}
which reaches 1 when the two quantities perfectly match.
The second metric adopted to evaluate the prediction accuracy is the Normalized Mean Square Error ($\nmse$), computed as
\begin{equation}\label{eq:nmse}
    \nmse
    \left( 
    \mathbf{\hat{v}}(\omega), \mathbf{v}(\omega)\right)
    = 10\log_{10} 
    \left(
    \frac{\norm[1]{\mathbf{\hat{v}}(\omega) - \mathbf{v}(\omega)}^2_{2} }
    {\norm[1]{\mathbf{v}(\omega)}^2_{2}} 
    \right).
\end{equation}
\begin{figure}[tb]
    \centering
    \includegraphics[width=0.47\textwidth]{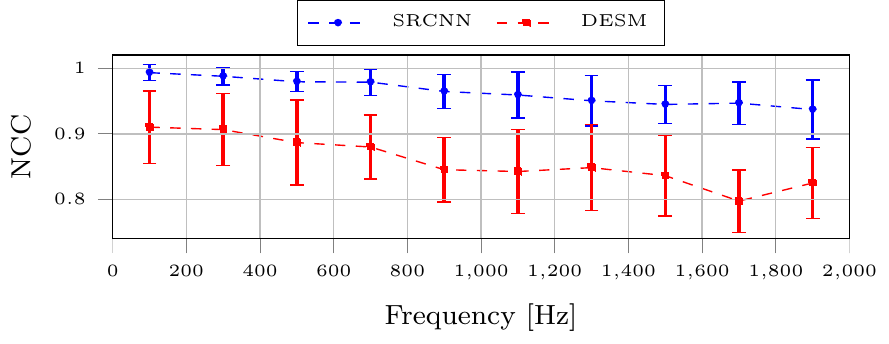}
    \vspace*{-1.5mm}
    \caption{NCC mean and standard deviation as function of frequency bands.}\label{fig:ncc_freq}\vspace{-0.5mm}
\end{figure}
\vspace*{-2mm}
\subsection{Network Validation}\label{subsec:validation}
The proposed SRCNN-NAH method was first evaluated with a noiseless version of the synthesized test set and validated in terms of $\nmse$ \eqref{eq:nmse} and $\ncc$ \eqref{eq:ncc}.
Samples have been sorted in ascending order of NMSE.

Inspecting Fig.~\ref{fig:metrics_cnn}(a) we can observe an accurate reconstruction for a wide range of vibrational images.
A decrease in $\nmse$ accuracy can be observed in Fig.~\ref{fig:metrics_cnn}(a); this involves mainly high frequency modes with complex patterns and it is caused by scaling errors between the network prediction and the reference values.
Nevertheless, by unfolding the definition used for the $\nmse$ presented in \eqref{eq:nmse}, we can infer that the average value $\nmse_{\mathrm{AVG}}=\SI{-10.96}{\decibel}$ corresponds to a relative error of $\SI{8}{\percent}$ related to a scaling bias.
The same behavior was also noted in \cite{olivieri2020nah}.

On the other hand, the average $\ncc$ value depicted in Fig.~\ref{fig:metrics_cnn}(b), $\ncc_{\mathrm{AVG}}=0.96 \,(\SI{96}{\percent})$, highlights the patter similarity for the majority of the test set with a minimum $\ncc$ value of $0.73\,(\SI{73}{\percent})$.

In order to evaluate the devised methodology with respect to state-of-the-art approaches, we compared the estimations obtained by SRCNN-NAH with the DESM approach of \cite{canclini2017dictionary}.
To the best of our knowledge, it is the first time that DESM is applied to anisotropic objects with arbitrary shapes. 
The equivalent sources are organized in a fixed regular rectangular grid of $20\times24$ points deployed in order to contain the projection of the violin plate with the largest dimensions  on the equivalent source plane. We refer the interested reader to \cite{canclini2017dictionary} for details on the setup.
The fictitious sources are placed $\SI{20}{\milli\meter}$ below the bottom surface of the violin plate.
The dictionary was built considering the velocity images related to the training set, then the technique was tested on the same test set of the SRCNN-NAH.

In Fig.~\ref{fig:ncc_freq} the $\ncc$ for both the DESM and the proposed SRCNN are reported as a function of the frequency. 
The mean and the standard deviation confidence error are computed considering the modal frequency in $10$ equally-spaced frequency bands in the range $[0, 2000]\,\si{\hertz}$ with frequency step size of $\SI{200}{\hertz}$.
We observe in Fig.~\ref{fig:ncc_freq} that SRCNN achieves an average $\ncc$ steadily above $0.9$ for the whole frequency range. 
Noteworthy, the standard deviation obtained by SRCNN is always smaller than the one reported by DESM.

\begin{figure}[tb]
    \centering
    \includegraphics[width=0.47\textwidth]{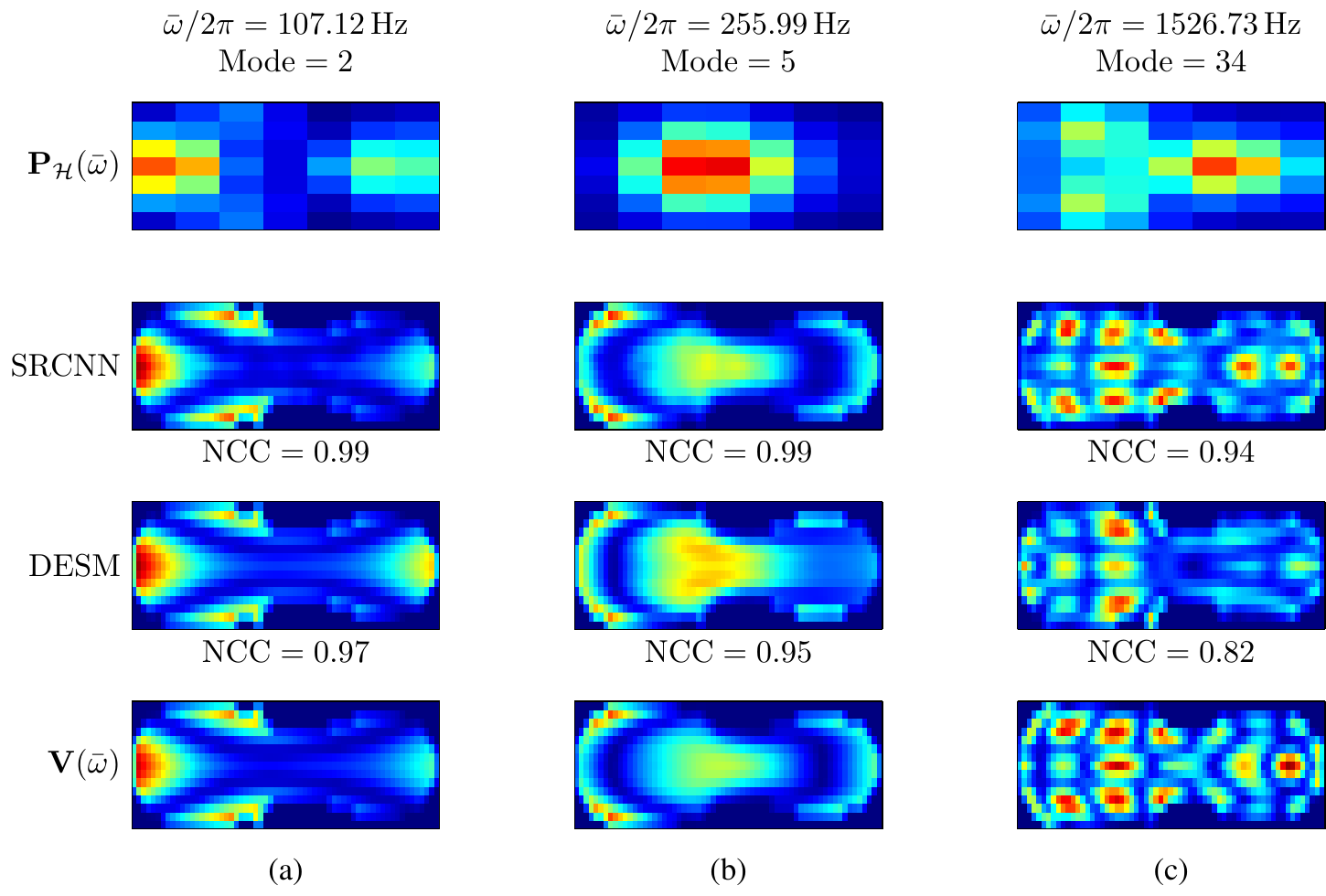}
    \vspace*{-1.5mm}
    \caption{Reconstruction examples of three different violin plates.}\label{fig:reconstructions}\vspace{-1mm}
\end{figure}
Three examples of estimations are reported in Fig.~\ref{fig:reconstructions}.
The hologram input data is in the first row.
The vibrational velocities predicted by SRCNN are depicted in the second row, while the DESM estimates and the ground truth are in the last two rows, respectively.
For the reconstruction of Fig.~\ref{fig:reconstructions}(a) and Fig.~\ref{fig:reconstructions}(b) both models retrieve the general shape of the velocity patterns.
Nevertheless, DESM shows higher scaling errors and lower NCC values than SRCNN.

\begin{figure}[tb]
    \centering
    \includegraphics[width=0.48\textwidth]{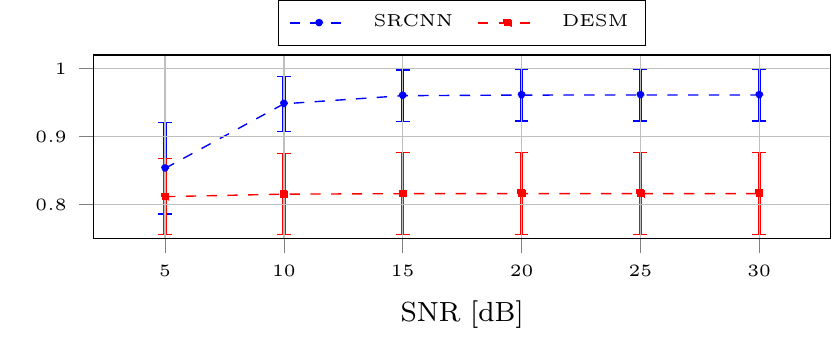}
    \vspace*{-1.5mm}
    \caption{NCC mean and standard deviation as function of $\mathrm{SNR}$.}\label{fig:ncc_snr}\vspace*{-3mm}
\end{figure}

Lastly, we investigated the robustness of the proposed SRCNN against noisy input data.
We tested the model with $6$ test sets corrupted by different realizations of additive white gaussian noise with varying $\mathrm{SNR} \in [5,30]\,\si{\decibel}$.
Fig.~\ref{fig:ncc_snr} shows the mean $\ncc$ value evaluated at different $\mathrm{SNR}$ along with the corresponding standard deviation.
As expected from results in \cite{olivieri2020nah}, the adoption of CNN provides robustness to noisy data, and SRCNN outperformed DESM.
In particular, SRCNN retrieves stable results down to a value of $\mathrm{SNR} = \SI{10}{\decibel}$, below which the performance decreases.

\section{Conclusions}\label{sec:conclusion}
In this paper, an enhanced CNN-based NAH solution has been proposed.
The devised architecture (SRCNN) enables the normal velocity estimation on orthotropic violin top plates starting from the pressure acquired by a limited number of microphones.

We validated the model predictions with respect to the velocity ground truth synthesized with FEM simulations.
Results showed a higher accuracy with respect to state-of-the-art NAH technique based on compressed sensing.
Moreover, the proposed method showed robustness against input data affected by additive white noise.
As future works, we will investigate on the use of physics-informed CNN architectures to provide better reconstruction accuracy. Moreover, we will analyse transfer learning techniques to solve the NAH problem to a set of complex-shape objects different from violins.

\bibliography{main.bib}

\begin{thebibliography}{10}

\bibitem{maynard1985nearfield}
J.~D. Maynard, E.~G. Williams, and Y.~Lee, ``Nearfield acoustic holography: I.
  theory of generalized holography and the development of {NAH},'' {\em The
  Journal of the Acoustical Society of America}, vol.~78, no.~4,
  pp.~1395--1413, 1985.

\bibitem{williams1999fourier}
E.~G. Williams, {\em Fourier acoustics: sound radiation and nearfield
  acoustical holography}.
\newblock Elsevier, 1999.

\bibitem{norton_karczub_2003}
M.~P. Norton and D.~G. Karczub, {\em Fundamentals of Noise and Vibration
  Analysis for Engineers}.
\newblock Cambridge University Press, 2~ed., 2003.

\bibitem{williams2001regularization}
E.~G. Williams, ``Regularization methods for near-field acoustical
  holography,'' {\em The Journal of the Acoustical Society of America},
  vol.~110, no.~4, pp.~1976--1988, 2001.

\bibitem{kim2004optimal}
Y.~Kim and P.~A. Nelson, ``Optimal regularisation for acoustic source
  reconstruction by inverse methods,'' {\em Journal of sound and vibration},
  vol.~275, no.~3-5, pp.~463--487, 2004.

\bibitem{scholte2008wavenumber}
R.~Scholte, I.~Lopez, N.~B. Roozen, and H.~Nijmeijer, ``Wavenumber domain
  regularization for near-field acoustic holography by means of modified filter
  functions and cut-off and slope iteration,'' {\em ACTA Acustica united with
  Acustica}, vol.~94, no.~3, pp.~339--348, 2008.

\bibitem{banerjee1981boundary}
P.~K. Banerjee and R.~Butterfield, {\em Boundary element methods in engineering
  science}, vol.~17.
\newblock McGraw-Hill London, 1981.

\bibitem{cheng2005bem}
A.~Cheng and D.~Cheng, ``Heritage and early history of the boundary element
  method,'' {\em Engineering Analysis with Boundary Elements}, vol.~29,
  pp.~268--302, 03 2005.

\bibitem{veronesi_digital_1989}
W.~A. Veronesi and J.~D. Maynard, ``Digital holographic reconstruction of
  sources with arbitrarily shaped surfaces,'' {\em The Journal of the
  Acoustical Society of America}, vol.~85, pp.~588--598, Feb. 1989.

\bibitem{schuhmacher2003ibem}
A.~Schuhmacher, J.~Hald, K.~Rasmussen, and P.~Hansen, ``Sound source
  reconstruction using inverse boundary element calculations,'' {\em The
  Journal of the Acoustical Society of America}, vol.~113, pp.~114--27, 02
  2003.

\bibitem{chardon2012near}
G.~Chardon, L.~Daudet, A.~Peillot, F.~Ollivier, N.~Bertin, and R.~Gribonval,
  ``Near-field acoustic holography using sparse regularization and compressive
  sampling principles,'' {\em The Journal of the Acoustical Society of
  America}, vol.~132, no.~3, pp.~1521--1534, 2012.

\bibitem{nachosURL}
G.~Chardon, L.~Daudet, A.~Peillot, F.~Ollivier, N.~Bertin, and R.~Gribonval,
  ``Nachos database and toolbox.'' \url{http://echange.inria.fr/nah/}, 2013.

\bibitem{koopman1989wavesuperposition}
G.~H. Koopmann, L.~Song, and J.~B. Fahnline, ``A method for computing acoustic
  fields based on the principle of wave superposition,'' {\em Journal of the
  Acoustical Society of America}, vol.~86, no.~6, pp.~2433--2438, 1989.

\bibitem{lee2016esmreview}
S.~Lee, ``Review: The use of equivalent source method in computational
  acoustics,'' {\em Journal of Computational Acoustics}, p.~1630001, 2016.

\bibitem{fernandez2017sparse}
E.~Fernandez-Grande, A.~Xenaki, and P.~Gerstoft, ``A sparse equivalent source
  method for near-field acoustic holography,'' {\em The Journal of the
  Acoustical Society of America}, vol.~141, no.~1, pp.~532--542, 2017.

\bibitem{ANTONI2019289}
J.~Antoni, J.~Le~Magueresse, Q.~Leclère, and P.~Simard, ``Sparse acoustical
  holography from iterated bayesian focusing,'' {\em Journal of Sound and
  Vibration}, vol.~446, pp.~289--325, 2019.

\bibitem{canclini2017dictionary}
A.~Canclini, M.~Varini, F.~Antonacci, and A.~Sarti, ``Dictionary-based
  equivalent source method for near-field acoustic holography,'' in {\em 2017
  IEEE International Conference on Acoustics, Speech and Signal Processing
  (ICASSP)}, pp.~166--170, IEEE, 2017.

\bibitem{Goodfellow2016Deeplearing}
I.~Goodfellow, Y.~Bengio, and A.~Courville, {\em Deep Learning}.
\newblock MIT Press, 2016.
\newblock \url{http://www.deeplearningbook.org}.

\bibitem{olivieri2020nah}
M.~Olivieri, M.~Pezzoli, R.~Malvermi, F.~Antonacci, and A.~Sarti, ``Near-field
  acoustic holography analysis with convolutional neural networks,'' {\em
  INTER-NOISE and NOISE-CON Congress and Conference Proceedings}, vol.~261,
  no.~1, pp.~5607--5618, 2020.

\bibitem{bainco2019acousticdeepreview}
M.~J. Bianco, P.~Gerstoft, J.~Traer, E.~Ozanich, M.~A. Roch, S.~Gannot, and
  C.-A. Deledalle, ``Machine learning in acoustics: Theory and applications,''
  {\em The Journal of the Acoustical Society of America}, vol.~146, no.~5,
  pp.~3590--3628, 2019.

\bibitem{he2016deep}
K.~He, X.~Zhang, S.~Ren, and J.~Sun, ``Deep residual learning for image
  recognition,'' in {\em Proceedings of the IEEE conference on computer vision
  and pattern recognition}, pp.~770--778, 2016.

\bibitem{campagnoli2020vibrational}
C.~Campagnoli, M.~Pezzoli, F.~Antonacci, and A.~Sarti, ``Vibrational modal
  shape interpolation through convolutional auto encoder,'' vol.~261, no.~1,
  pp.~5619--5626, 2020.

\bibitem{acerbi2021interpolation}
M.~Acerbi, R.~Malvermi, M.~{Pezzoli}, F.~{Antonacci}, A.~{Sarti}, and
  R.~Corradi, ``Interpolation of irregularly sampled frequency response
  functions using convolutional neural networks,'' in {\em International
  Conference on Acoustics, Speech and Signal Processing, {(ICASSP)}}, IEEE,
  2021.

\bibitem{rawat2017cnnimagereview}
W.~Rawat and Z.~Wang, ``Deep convolutional neural networks for image
  classification: A comprehensive review,'' {\em Neural Computation}, vol.~29,
  pp.~1--98, 2017.

\bibitem{ronneberger2015unet}
O.~Ronneberger, P.~Fischer, and T.~Brox, ``U-net: Convolutional networks for
  biomedical image segmentation,'' in {\em International Conference on Medical
  image computing and computer-assisted intervention}, pp.~234--241, Springer,
  2015.

\bibitem{nair2010rectified}
V.~Nair and G.~E. Hinton, ``Rectified linear units improve restricted boltzmann
  machines,'' in {\em Proceedings of the 27th international conference on
  machine learning (ICML-10)}, pp.~807--814, 2010.

\bibitem{kingma2014adam}
D.~P. Kingma and J.~Ba, ``Adam: A method for stochastic optimization,'' {\em
  arXiv preprint arXiv:1412.6980}, 2014.

\bibitem{chollet2015keras}
F.~Chollet {\em et~al.}, ``Keras.'' \url{https://keras.io}, 2015.

\bibitem{gonzalez2021data}
S.~Gonzalez, D.~Salvi, D.~Baeza, F.~Antonacci, and A.~Sarti, ``A data-driven
  approach to violin making,'' {\em Scientific Reports}, vol.~11, no.~1,
  pp.~1--9, 2021.

\bibitem{salvi2021parametric}
D.~Salvi, S.~Gonzalez, F.~Antonacci, and A.~Sarti, ``Parametric optimization of
  violin top plates using machine learning,'' {\em arXiv preprint
  arXiv:2102.07133}, 2021.

\bibitem{comsol2018v54}
S.~COMSOL~AB, Stockholm, ``Comsol multiphysics v. 5.4.''
  \url{https://www.comsol.com}, 2018.

\bibitem{ross2010wood}
R.~J. Ross {\em et~al.}, ``Wood handbook: wood as an engineering material,''
  {\em USDA Forest Service, Forest Products Laboratory, General Technical
  Report FPL-GTR-190, 2010: 509 p. 1 v.}, vol.~190, 2010.

\end{thebibliography}
\bibliographystyle{ieeetr}
\end{document}